# Space Public Outreach Team: Successful STEM Engagement on Complex Technical Topics

ANGELA DES JARDINS
*Montana State University, USA*
angela.desjardins@montana.edu

JOEY SHAPIRO KEY
*University of Washington Bothell, USA*
joeykey@uw.edu

KATHRYN WILLIAMSON
*West Virginia University, USA*
kewilliamson@mail.wvu.edu

SETH KIMBRELL
*Montana State University, USA*
sethkimbrell@montana.edu

SOPHIE DE SAINT-GEORGES
*Green Bank Observatory, USA*
Sdesaint@nrao.edu

TYSON LITTENBERG
*NASA Marshall Space Flight Center, USA*
Tyson.b.littenberg@nasa.gov

JESSICA PAGE
*University of Alabama Huntsville, USA*
Jp0089@uah.edu

TIMOTHY DOLCH
*Hillsdale College, USA*
tdolch@hillsdale.edu

It is the responsibility of today's scientists, engineers, and educators to inspire and encourage our youth into technical careers that benefit our society. Too often, however, this responsibility is buried beneath daily job demands and the routines of teaching. Space Public Outreach Team (SPOT) programs leverage a train-the-trainer model to empower college students to make meaningful impacts in their local communities by engaging and inspiring younger students through science presentations. SPOT takes advantage of the excitement of space and the natural way college students serve as role models for children. The result is a win-win program for all involved. This paper describes the original Montana SPOT program, presents analyses demonstrating the success of SPOT, gives overviews of program adaptations in West Virginia and with the NANOGrav collaboration, describes how college student presenters are able to share complex topics, and discusses the importance of college student role models. We hope that our experiences with SPOT will help others implement similar strategies in their own communities.

***Keywords:*** STEM education, outreach, space science

## INTRODUCTION

The Space Public Outreach Team (SPOT) began in 1996 as a program of the Montana Space Grant Consortium (MSGC). MSGC is a member of a national network of 52 NASA-funded consortia, working to strengthen aerospace education and research in the United States. The primary goal of SPOT is to inspire youth to engage in science, technology, engineering, and math (STEM), either as a career or in order to promote general technical literacy. To accomplish this goal, SPOT takes advantage of both the excitement of space and the natural way college students serve as role models for children. The basic SPOT model is to send specially-trained college students into K-12 classrooms to share current space science topics via interactive presentations. The result is a win-win program for the college student presenters who gain valuable experience, the K-12 students and educators, the host organization, and NASA.

The basic SPOT model, training college students to share space topics with young people, has been quite successful. This paper provides an analysis supporting this claim. First, a description of how the model has been formally adopted in several locations is presented. Next, empirical evidence of interest and impact for the major program components is addressed. In this paper, "undergraduates", "college students", "ambassadors", and "presenters" all refer to the students who are giving presentations to the K-12 schools and other programs.

**Need for the SPOT program**

There is a need for programs like SPOT that inspire young people to pursue STEM careers. The President's Council of Advisors on Science and Technology (PCAST, 2012) report *Engage to excel: producing one million additional college graduates with degrees in science, technology, engineering, and mathematics (STEM)* stated that fewer than 40% of students who enter college intend to complete a STEM degree, and even less complete the degree. In order to fill the widening gap for STEM professionals in our nation's economy, we need to increase that number to more than 50%. The PCAST report also described the need to double our efforts to engage underrepresented groups in STEM fields. In addition, the report warned that better teaching methods are needed to make courses more inspiring and to create an atmosphere of a community of STEM learners.

The SPOT program serves the need to recruit more students into STEM fields by working to create a community of learners teaching younger learners, or near-peer mentoring. Near-peer mentoring is effective for sustainable science education outreach, both because near-peers help the younger students envision themselves as future scientists (Markus & Nurius, 1986) and because college students are more available than professors or subject matter experts (Pluth, Boettcher, Navin, Greenway, & Hartle, 2015). SPOT takes advantage of natural curiosity about space to engage K-12 students in the discovery of science, engineering, and possible STEM careers.

Near-peer interactions are particularly important for underrepresented groups (Trujillo et al., 2015). All SPOT programs have a special focus on reaching minority, low income, and rural students. The SPOT program focuses on these groups by advertising the program opportunity in schools with higher percentages of underrepresented students. In addition, the program works to recruit presenters who are themselves underrepresented, encouraging every presenter to spend a bit of time introducing themselves so that the K-12 students can better identify with them and ask interesting questions. Finally, SPOT tries to match schools with presenters who are as geographically close as possible. It is much more meaningful when a presenter is from a nearby college or university.

SPOT presenters are also learners, as they gain valuable experience in communicating science. There is broad agreement that the science community would benefit from additional science communication training (Besley & Tanner, 2011). While graduate student experts create the SPOT presentations, the SPOT presenters spend significant time learning the presentation and the presentation subject. After giving a successful "practice presentation" to program leadership, the presenters then turn their new knowledge and skills directly into teaching. In the process of giving SPOT presentations, the college students learn leadership skills, how to speak well, and how to communicate science effectively to a diverse audience.

**Paper Overview**

Information about the currently active SPOT programs is first provided in the background section of this paper. Next is a description of methods: why the SPOT model is successful, the power of the SPOT presenter role model, and how complex technical topics are shared. Following the methods section is a discussion of assessment of SPOT's impact and program challenges. Finally, a summary and list of future prospects is provided.

## BACKGROUND

SPOT began as the Mars Pathfinder Outreach Project (MPOP) in the fall of 1996. After the success of the Pathfinder landing in 1997, it became clear NASA was going to have an "every two years to Mars" plan, and MPOP morphed into the Mars *Public* Outreach Project. The following academic year, the program was generalized to be the *Space* Public Outreach Team and Montana adopted the present method of focusing each year's show on a current space topic. This fits well with Montana Space Grant Consortium's (MSGC) mission to share NASA's unique content with Montana communities. When West Virginia adopted SPOT, they expanded the scope further; their program was named the *Science* Public Outreach Team. The North American Nanohertz Observatory for Gravitational Waves (NANOGrav) SPOT program, on the other hand, always gives presentations focused on gravitational waves and their detection. These adaptations are discussed further in the sections below.

### The Original Montana SPOT Model

The organization of SPOT involves students from kindergarten through graduate school as well as educators and institution staff. Institution staff and faculty members oversee the program financially, recruit and train the managers, select presentation topics, and generally provide a connection between the institution and the SPOT program. SPOT graduate student managers recruit and train the undergraduate college student presenters. In addition, the managers are responsible for coordinating visits between the K-12 educators and presenters, travel-related arrangements, creating new presentations, and helping collect reporting and evaluation data. The college students learn the presentation, and once they have passed their practice session with the SPOT graduate student manager, choose to take on slots that fit their schedule. The presenters are paid for their travel and presentation time. On the receiving end, SPOT engages young students in their classrooms or as part of an out-of-school or community program such as Girl Scouts. Teachers or group leaders are encouraged to spend additional classroom or meeting time interacting with the students on the SPOT presentation topic. To aid the teachers, online materials as well as printed materials are provided in a "teacher pack".

The Montana SPOT program reaches between 8,000 and 15,000 young people each year. While the program aims to reach high numbers of students, the program also has a focus on engaging rural and native populations, who are especially in need of external classroom visits. Upon completion of visits to rural schools, the following remark is often heard, "Great, we got our science for the year!" Comments like this are discouraging but point to the importance of programs like SPOT.

The key factor in being able to carry out a large number of visits is recruiting a large pool of presenters. Recruiting presenters is somewhat complex. Success depends on the advertising method, the style of the managers, and how the program is presented in the initial informational meeting. Presently, the Montana program is advertised directly to K-12 educators in accordance with the number of certified (i.e. passed their practice presentation) presenters. It is important to balance the number of requests that can be accommodated with the availability of the certified presenters.

SPOT recruits presenters from as many Montana colleges and universities as possible. Having presenters in a variety of locations across the state helps the program reach more diverse populations for lower cost due to reduced travel time. Additionally, it is more impactful to have a presenter from the local campus visit a rural classroom, versus a student from a university in a far-away city. Figure 1 shows a SPOT presenter from Rocky

Mountain College presenting to a small school on the Crow Reservation in 2018. In the past, training students in remote locations was very difficult and required separate managers or staff support at each site. With video conferencing more readily available, however, training and working with remote presenters is now easier.

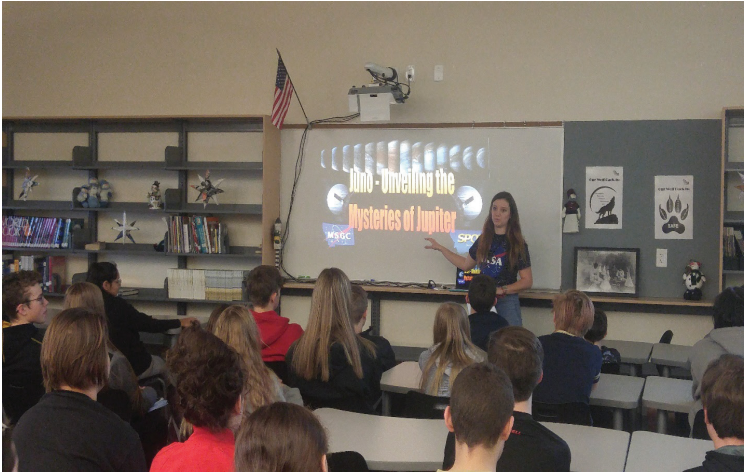

**Figure 1.** Montana student presenting to a small school on the Crow reservation in 2018.

From 2004-2008, the program received additional funding as part of a NASA education and public outreach effort. With this funding, the program began a focus on reaching students in rural areas and areas that have high percentages of Native American students. At the completion of this expansion time, we conducted an evaluation of the program. Figure 2 shows the reach of the SPOT program across Montana for the five-year period 2004-2008. During these five years, Montana Space Grant SPOT reached approximately 55,000 students, several of whom went on to be SPOT presenters in college.

Montana is the nation's fourth largest state geographically. The program endeavors to reach all parts of the state but it is very difficult when, for example, driving from Montana State University in Bozeman, MT to Plentywood High School takes 7.5 hours in each direction. One way the SPOT program deals with the large travel distances is to rotate to different remote locations each year, rather than presenting at the same remote school every year. Another strategy is to not accept a presentation request that requires an overnight stay unless the presentation is for a minimum of 100 students. To accommodate student presenters' lack of time for long trips, remote presentations are typically scheduled during opportune times such as early January (college students are still on break but the K-12 students are not), during the university spring break (earlier than most of the K-12 school breaks), and during the second half of May (college is out but K-12 schools are not).

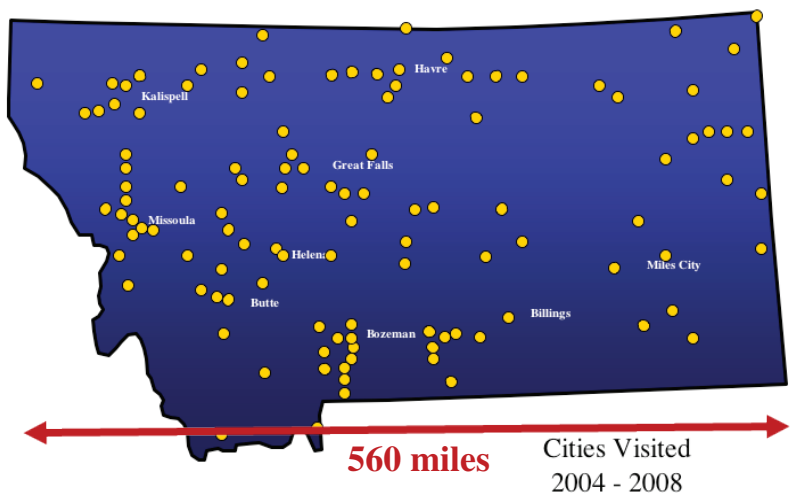

**Figure 2.** Example of statewide reach in Montana, 2004-2008.

### Adaptation of the Model to West Virginia

In 2013, the Green Bank Observatory (GBO) adapted the SPOT model for rural West Virginia (Williamson, 2013; Williamson et al., 2014). The GBO houses some of the world's largest and most advanced radio telescopes. Paradoxically, the presence of these advanced instruments requires residents in the surrounding communities to live without common technologies like WiFi or cell phones because they interfere with radio observations. While this hinders residents' access to some modern technology and education, the facilities, instruments, and personnel at the GBO provide a powerful local educational outreach opportunity for WV SPOT to utilize.

Since its inception in 2013, WV SPOT has trained 181 ambassadors (WV refers to their presenters as ambassadors) at seven West Virginia col-leges and universities. One hundred thirty-two (132) of these ambassadors (73%) went on to give 765 presentations at 299 schools (i.e. ranging from 46-194 presentations and 27-80 schools per year) for over 24,000 K-12 students and 1,100 teachers (i.e. 2,800-5,000 people each year).

The WV SPOT program is directed by education officers at the Green Bank Observatory (GBO) and each participating college campus has a faculty advisor. During the fall of each academic year, campus advisors recruit college ambassadors to apply for travel to GBO for an immersive training weekend in the National Radio Quiet Zone (Figure 3). The training weekend is focused on science communication, team building, and learning the skills such as how to run the electronic polling software. Ambassadors practice the featured SPOT presentations and hands-on activities, and stock their campus kits with hands-on activity supplies.

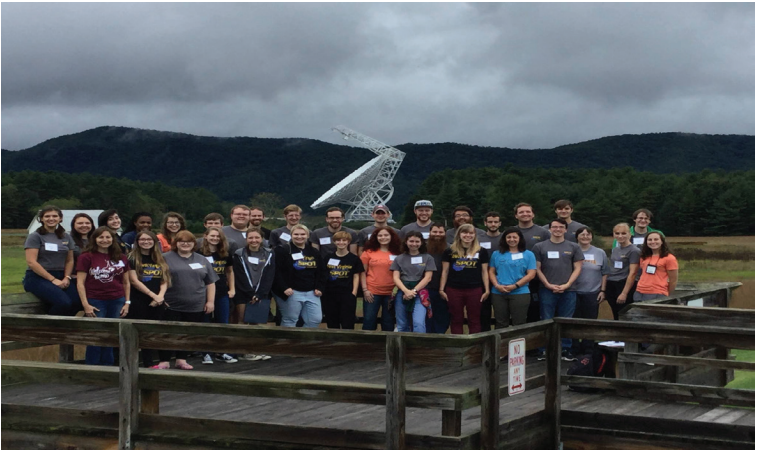

**Figure 3.** West Virginia SPOT Ambassador training at the Green Bank Telescope in 2018.

An important addition of the WV SPOT model is a hands-on activity component. There are currently nine hands-on activities offered that teachers can choose to accompany a presentation, though the maximum class size that can be accommodated is 30. The activities require minimal, low-cost supplies that can be obtained from any local grocery store, and they can significantly enhance the slide show presentation. For example, in the "Green Bank Telescope Engineering Design Challenge," students use gumdrops, tooth picks, marshmallows, and spaghetti to design and build a tall structure that can hold weight, gaining first-hand experience for what professional engineers must consider when designing new telescopes. Lesson plans and material lists that go with each hands-on activity are provided on the WV SPOT website (www.wvspot.org).

Operation of the WV SPOT program relies heavily on in-kind contributions of GBO management and campus faculty advisors' time, but the primary direct costs are ambassador trainings, honoraria, and travel. A significant fraction of these costs are provided through partnership with the West Virginia Space Grant Consortium. As of 2016, SPOT also created sponsorship and membership policies to build a more sustainable funding mechanism, which has proven to be extremely successful. Companies and businesses are invited to sponsor SPOT through a donation, receiving advertising via their logos on presentations. Researchers are invited to become WV SPOT members by developing a presentation that features their research, allowing them to plug-in to the SPOT model and accomplish broader impacts for their grants. So far, two companies have joined as sponsors and four research teams have joined as members to showcase topics such as pulsars and fast radio bursts, plasma and space weather, and water quality research and sustainability. A podcast episode that follows WVU ambassadors leading a water resources presentation was created to share with other presenters (Mazzella, 2017). WV SPOT membership is now featured through the West Virginia University Research Office and is generating significant interest as a promising way for researchers to collaborate across disciplines to support education and outreach.

### Adaptation of the Model to NANOGrav

The North American Nanohertz Observatory for Gravitational waves (NANOGrav) is a distributed scientific collaboration with scientists and students at over 30 institutions across the US and Canada. The SPOT model has been adapted to fit the distributed NANOGrav collaboration by coordinating with the NANOGrav Student Team of Astrophysics ResearcherS (STARS) program and leveraging connection to students through the NANOGrav senior researchers. The NANOGrav STARS program trains undergraduate students to search for pulsars in data from radio telescopes and contribute to the search for gravitational waves in the NANOGrav pulsar timing array. STARS nodes at NANOGrav institutions are led by a faculty researcher and Team Leader students and are connected through weekly telecons that include research project updates and professional development sessions. STARS students are encouraged to become SPOT Ambassadors (NANOGrav also refers to presenters as ambassadors) to support their research training, including learning the basics of gravitational wave astron-omy and pulsar timing arrays while developing their public speaking skills.

The NANOGrav SPOT program is managed by a graduate student in collaboration with senior researchers. Each participating institution has a point of contact who communicates with the program manager and local students and teachers. Requests for SPOT presentations come through a request form on the NANOGrav website as well as through direct contact with local schools and community groups. Presenters enter reporting information with a single form on the NANOGrav website used to collect data for the distributed program.

The benefits of a distributed SPOT program include a broad geographic and demographic reach for presenters and audience members. For example, Figure 4 shows the NANOGrav presentation given in a Milwaukee classroom and Figure 5 shows the same presentation given to a general public audience at the Vanderbilt Dryer Observatory in Tennessee.

Figures 4 and 5 also demonstrate the large range of educational technologies supporting SPOT presentations. Specifically, Figure 4 features a simple display screen for a small gathering of students in their classroom, while Figure 5 shows a presentation underneath replicas of a satellite and receiver dish, to a sizeable group. SPOT presenters need to be comfortable with adjusting on short notice to the available technologies on site.

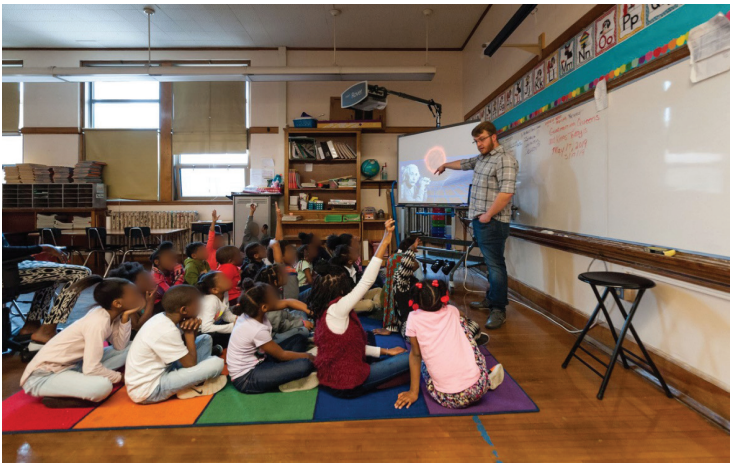

**Figure 4.** A student from the University of Wisconsin Madison presents the NANOGrav show in a Milwaukee classroom.

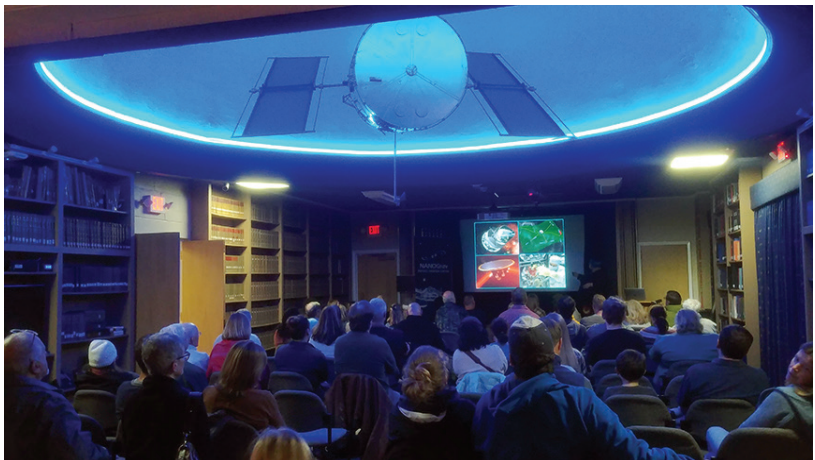

**Figure 5.** General public audience at a NANOGrav SPOT presentation at a "Meet the Astronomer Night" event at Vanderbilt Dyer Observatory.

Undergraduate students at institutions without similar outreach programs can engage with the NANOGrav network to contribute to and benefit from the distributed collaboration. It can be challenging to manage and communicate across institutions, including recruiting and training new presenters and establishing local knowledge of the program. The active NANOGrav SPOT presenters form a loose cohort but do not have the opportunity to meet in person except with the other presenters at their own institution. The program connects students across institutions but would benefit from increased student interaction.

## METHODS

### Why the SPOT Model is Successful

There is evidence that the SPOT program is successful through quantitative and qualitative data collected over the last 24 years from students, educators, presenters, managers, and program sponsors. As an example, Figure 6 shows NANOGrav student ambassador evaluations of their 151 SPOT visits from 2015-2019. Ratings are scaled from 1-5, where 5 is strongly agree, estimating the quality of students' interest and engagement in the presentation, as well as the success in presenting material to students at the appropriate age level. Individual presentations are customized to the location and age group of the audience. These data indicate that students are engaged and the presentations are geared at the appropriate level for the audience.

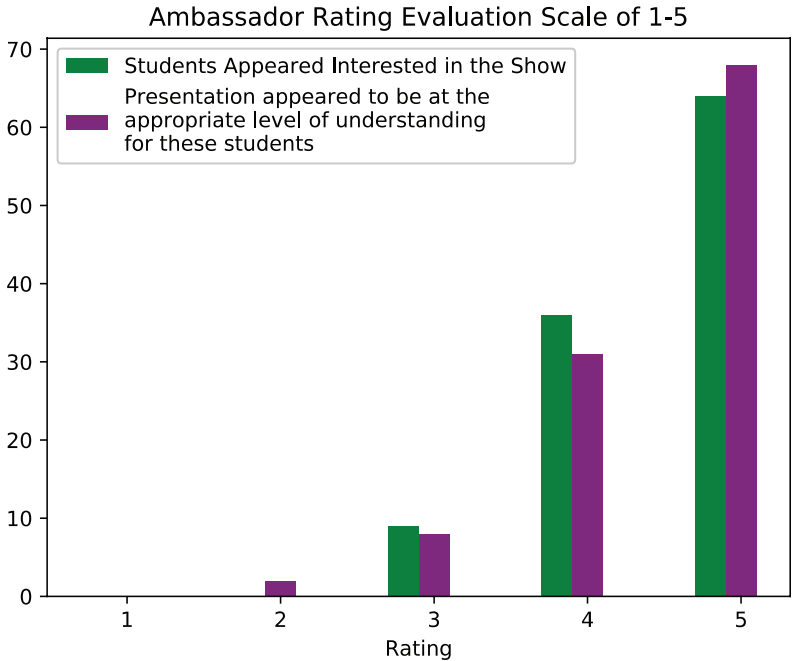

**Figure 6.** NANOGrav ambassador evaluations of the 151 SPOT visits, 2015-2019.

Another example indicating success is that SPOT is invited back to classrooms year after year. Sometimes presenters receive a packet full of colorful thank you notes from the students (Figure 7)! Additional information about how each program evaluates effectiveness is included in the following sections.

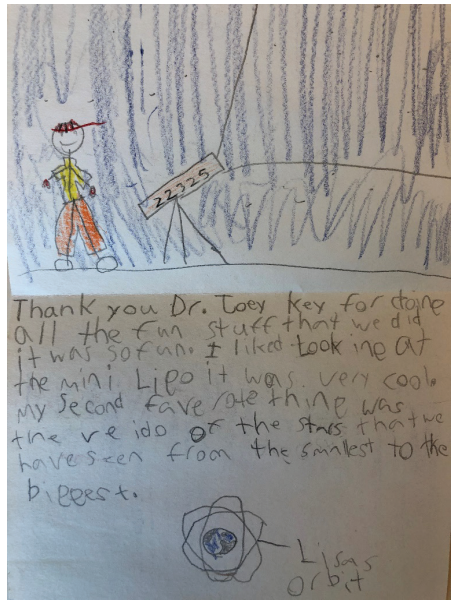

**Figure 7.** Example thank you note received from a NANOGrav SPOT presentation.

*Benefits of Team Effort.* One outstanding feature of the SPOT model is that it is both educational and enjoyable to all parties involved. The graduate student managers can improve their public speaking skills, as well as gain experience leading and managing a team. For many, SPOT is their first experience managing people, and having on-the-job experience with a support system behind them is very beneficial. Additionally, the managers are supported on fellowship funding, allowing them to focus on SPOT as well as their regular studies. College student presenters receive hands-on speaking training presentations from local communication experts as well as links to training resources such as those provided by the American Astronomical Society Astronomy Ambassadors program. Presenters also learn about the science from the notes in the PowerPoint document created by the subject matter experts and through in-person presentations from local scientists. Participating in SPOT gives confidence, develops leadership abilities, and helps presenters become comfortable with the technologies associated with showing material in an interactive way. Presenters appreciate helping others. One presenter who is now a teacher commented that SPOT helped her address her fears of teaching science; now she is excited to explain science topics to her students. Serving as a SPOT presenter is also an important way for students to have their resumes stand out, and can be featured in letters of recommendation. Paying presenters is important. While some students have the ability to volunteer their time, offering pay enables SPOT to recruit low income students who need to work.

Importance of the Presentations. Significant effort is put into the creation of the presentations each year. Design considerations facilitate the production and delivery of presentations that contain high-quality content but are tailored to local context needs and preferences of the presenters. Essential design considerations include the need to:

1. Provide accurate information while creating an engaging and relevant content.
2. Tie all the ideas together by creating a deliberate story line.
3. Include questions that can be answered by a simple raise of hands or through feedback technologies like clickers or Plickers.
4. Use the best available technology to show high quality videos and model simulations.
5. Have clean, simple visuals – SPOT presentations have almost no text. Detailed information for the presenter is added in the notes section of the PowerPoint slides.
6. Encourage presenters to spend more time on areas of special interest to them, to bring out their own personalities and excitement for selected topics.

Supplemental Learning Materials. In Montana, handouts are provided for each student to take home. Sometimes the printed materials are provided by NASA but often the team creates a postcard-sized handout specific to the current presentation. The postcard has engaging pictures on the front and information and links on the back. Figure 8 shows the front and back of the card created for a show about Jupiter and the Juno mission for the 2017-2018 academic year. These materials encourage continuing conversations with family members at home. All SPOT programs also take advantage of technology available on the web by providing links to online resources and opportunities that are related to the presented subject matter.

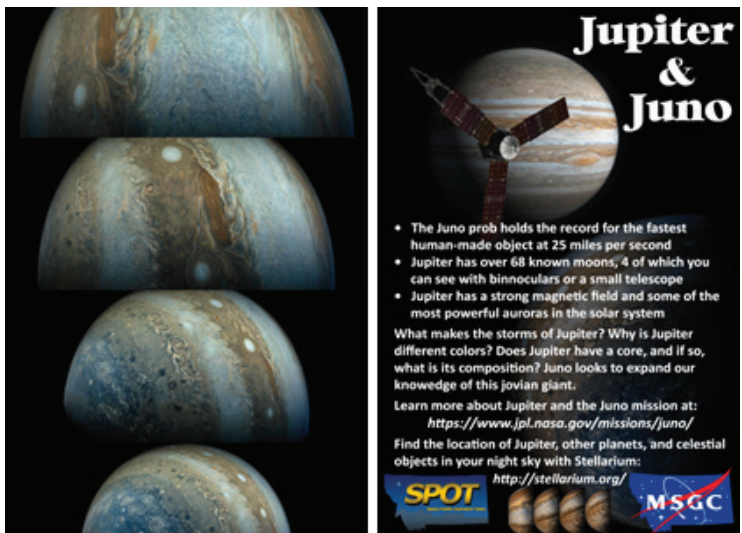

**Figure 8.** Front (left) and back (right) of the postcard given to each Montana student who saw the "Juno: Mission to Jupiter" presentation during the 2017-2018 academic year.

Low Cost to Recipient. SPOT presentations are offered at little (West Virginia) or no (Montana, NANOGrav) cost to schools and programs. The project's philosophy is that cost should not be a barrier to having access to this educational and engaging asset. Charging a substantial fee could severely reduce the number of presentations to poorer schools that need the interaction the most.

**Power of the SPOT Presenter Role Model**

A key component of the SPOT program is the role-model nature of the presenters. Presenters give an inspirational and relatable example to K-12 students interested in STEM. Presenters introduce themselves and their major at the beginning of each show, and students are encouraged to ask presenters questions relating to life in higher education or science in addition to content from the presentation. The 9$^{th}$ - 12$^{th}$ grade audience is often interested in how the presenter achieved their position and what courses or majors they recommend they pursue. The profound impact presenters leave is evident, considering the enthusiasm received by the audience during the visit and the wide range of questions asked, as well as follow-up letters, emails and requests for subsequent visits. The role models are especially helpful to schools in communities that lack frequent exposure to professionals in STEM. Handouts with links for further reading on the material mentioned are left with the students, and teachers are given the presenter's contact information for students' further questions.

Each SPOT presentation includes information on related research being done in the local area. It is important to convey to the young people that it is not necessary for them to travel large distances or go to an Ivy League school to do great science because researchers are working on exciting topics near their homes. Many students have never heard of these opportunities before the presentation. SPOT hopes to inspire some youth to pursue science and engineering careers.

**How SPOT Shares Complex Technical Topics**

SPOT shares topics that are readily engaging, but often explaining these technical topics can be challenging. Some topics can be difficult for adults as well as children, such as how one describes black holes and gravitational waves. By way of general guidance, SPOT presenters are encouraged to:
1. Tailor the presentation to the age of their audience.
2. Build background information into the presentation so the students understand the big picture of the topic.
3. Share relevant discoveries that help tell a story, rather than simply reciting facts.
4. Make analogies and comparisons to common experiences, such as what size the Earth would be if Jupiter were the size of a beach ball.
5. Incorporate some form of activity for the audience, such as having the students in the audience stand up and act out moons orbiting planets.
6. Use available local resources such as projectors and SMART boards to share the presentation.

These actions help presenters keep the audience more focused than they would be for a more lecture-style presentation.

SPOT programs use a wide variety of educational technologies to share complex topics and manage the program. For example, to assess audience learning, WV SPOT is beginning to experiment with a new technology called "Plicker," which stands for "paper clicker." Each audience member receives a paper with a unique black and white box pattern, which they can rotate into different orientations for different votes on multiple choice ques-tions (ex: A-D for the four edges of the card). The presenter uses a mobile app to scan the votes, and then all results are stored online, allowing for both pre-post comparisons and across-site comparisons of student learning. The SPOT presentation for NANOGrav, called "Tuning in to Einstein's Universe," includes stimulating videos of cutting edge computation models of black holes merging. To help address some of the management difficulties in coordinating logistics for presenter visits to schools, computer science students at Fairmont State University are developing a WV SPOT mobile application. With the app, presenters will be able to see the drive time to a school, filter by presentation, claim a request, and even communicate with other presenters/ambassadors.

## ASSESSMENT OF IMPACT

Montana SPOT program evaluation and data collection includes presenter reporting forms, educator comment forms, data from presentation requests, and occasional external professional evaluation. Data from the presenter forms is submitted to NASA in MSGC's annual reporting. Educator comments are analyzed to ensure the program is providing the best experience possible. Based on feedback from the presenters and educators, the presentation is updated. External evaluators have been hired to ascertain the overall effectiveness of the SPOT program. The resulting information was used to make changes in the operation, including creating special, simplified presentations for the youngest audiences and working to make the presentations as hands-on as possible. In addition to these more formal types of evaluation, the team members are continually thinking about ways to improve the program. Planned for the near future is to conduct pre and post survey of impacts on the K-12 student recipients. This type of study is very valuable but also difficult to carry out due to the involvement of minors as research participants.

Assessment of West Virginia SPOT has primarily been accomplished via (1) an online reporting form submitted by ambassadors when they return from a trip to a school, which includes audience numbers and presenter impressions of how the presentation went, and (2) a teacher evaluation form, which teachers fill out after the presentation, including their impressions of student engagement. Figure 9 shows West Virginia SPOT 2018-2019 annual evaluation from their presenters. From the 80 responses, nearly all agree or strongly agree that the K-12 students are engaged by the SPOT presentations. Figure 10 shows the most recent West Virginia SPOT annual evaluation from teachers. The presentations were given to 57 schools and orga-nizations with evaluations returned from 56 teachers for a 98% evaluation response rate. As seen in the chart, the majority of teachers strongly agree that they would recommend the program and that the presentation was relevant to their curriculum.

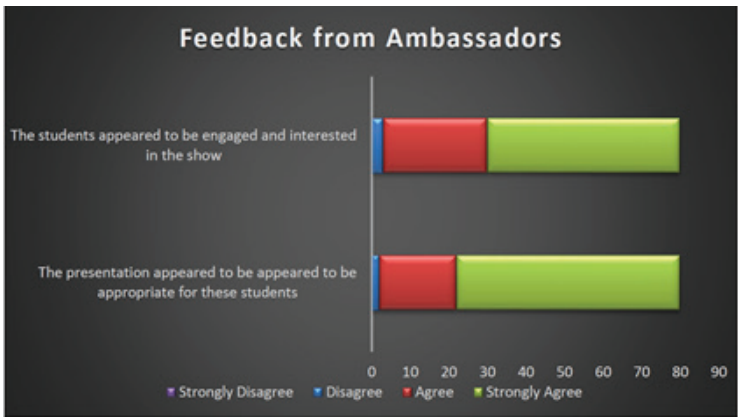

**Figure 9.** West Virginia SPOT 2018-2019 annual evaluation from presenters (Ambassadors).

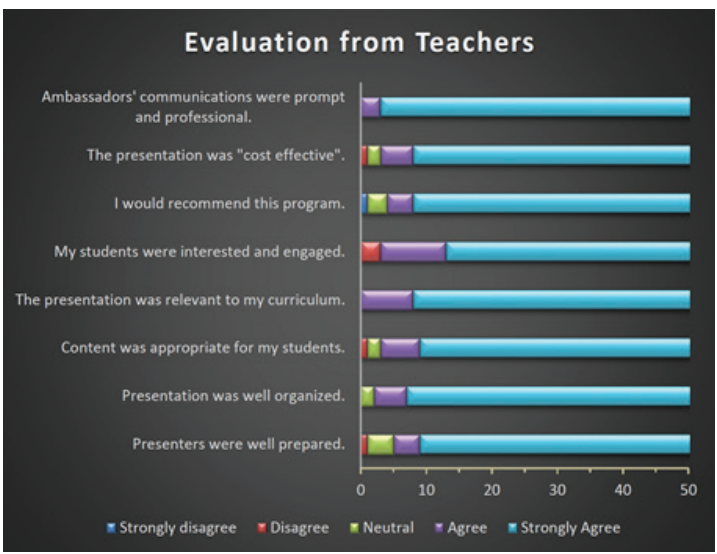

**Figure 10.** West Virginia SPOT annual evaluation from teachers 2018-2019.

Data on the NANOGrav adaptation of SPOT has been recorded since the fall semester of 2015. From 2015-2019, NANOGrav SPOT Ambassadors gave a total of 150 presentations, reaching 8,792 ± 1127 students in 10 states and Puerto Rico. The number of presentations and students reached in a given year steadily increased from prior years indicating the program will continue to grow. Post-presentation evaluations have shown positive results for both the presenters and the target audiences (see Figure 6). The SPOT audiences have reported a stronger desire for future STEM careers, as well as a better understanding of astronomy. On the presenter side, the presenting students have given positive feedback about their increased "soft skills" such as working in a team and science communication.

## DISCUSSION

While the general SPOT model is quite successful, there are several aspects that make carrying out the program challenging. Challenging aspects include motivating busy undergraduate student presenters to stay active, obtaining funding, covering large distances, and engaging young people on complex topics. West Virginia has done a good job addressing motivating their college student presenters. Each SPOT program has found its own method of sustaining funding. Montana offers the program for free and covers costs primarily under MSGC, though funding has come from other NASA entities in the past. West Virginia SPOT charges schools a small fee and is creative about finding other funding sources. The NANOGrav SPOT program has fewer overhead organizational costs and is supported by NANOGrav education and public outreach efforts and NSF NANOGrav Physics Frontiers Center (PFC) funding.

## FUTURE PROSPECTS

Beginning in 2020, Montana SPOT plans to test a new approach. Instead of focusing primarily on classroom presentations, some evening presentations with whole families invited will be scheduled. As an incentive for attendance, the team will work with the school's parent-teacher council to provide pizza. Research (Kreider, Caspe, Kennedy, & Weiss, 2007) shows that it is vitally important to a student's success to have at least one parental figure actively encouraging them. Additionally, when parents are involved in a young person's learning and cultivation of curiosity, the chance of student achievement rises dramatically. West Virginia SPOT has already conducted some presentations in this format and found it impactful.

Several other state Space Grant Consortia are considering creating SPOT programs. These include in Georgia, where a pilot program has already begun, in Oregon, in partnership with the Oregon Museum of Science and Industry (OMSI), and in Nevada, where the focus is on reaching their Native and Hispanic K-12 populations. In addition, plans are in place to include SPOT as an Education and Public Outreach (EPO) effort of the NASA Laser Interferometer and Space Antenna (LISA) mission. Finally, the NASA Science Mission Directorate Science Activation program NASA Space Science Education Consortium (NSSEC) is working to help share the SPOT model.

## SUMMARY AND CONCLUSIONS

In summary, the SPOT program is a successful activity for engaging students from kindergarten through graduate school. College students serve as role models for young people while also gaining valuable skills themselves. Graduate student managers provide a key link between the science, the college students, the K-12 educators, and the young people. The model was developed to enable creative engagement on complex topics. Another advantage of the model is that it is completely scalable and transferrable. The program is valuable to the funding entities, sharing exciting results and inspiring our youth to consider STEM careers. Ultimately, the power of SPOT comes from the natural way young people look up to college students and the ability of space to ignite a sense of wonder. The remainder is simply carefully thought out details.